\title {Detrimental Decoherence}
\author {  Gil Kalai\footnote{ 
Research supported in part by an NSF grant, 
an ISF grant, and a BSF grant. This paper is a revision and an extension 
of the formal part of \cite {Ka2}. 
I am grateful 
to Dorit Aharonov, Michael Ben-Or, Greg Kuperberg,  
and Robert Raussendorf
for 
fruitful discussions, 
and to many colleagues for helpful comments.
}
 \\
Hebrew University of Jerusalem and Yale University}
\documentstyle[12pt,amsfonts]{article}
\newtheorem{theo}{Theorem}

\newtheorem{lemma}[theo]{Lemma}

\newcommand{\beq}[1]{\begin{equation}\label{#1}}
\newcommand{\enq}[0]{\end{equation}}

\renewcommand{\baselinestretch}{1.21}

\begin{document}

\maketitle

\begin {abstract}
We propose and discuss two conjectures on the nature of information leaks 
(decoherence)
for quantum computers.
These conjectures, if (or when) they hold, are damaging 
for quantum error-correction 
as required by fault-tolerant quantum computation.  

The first conjecture asserts that information leaks for a pair of  
substantially entangled qubits are themselves substantially positively 
correlated. 

The second conjecture asserts that 
in a noisy quantum computer with highly entangled qubits there will 
be a strong effect of error synchronization. 

We present more general conjectures for arbitrary noisy
quantum systems.


\bigskip

\bigskip

\end {abstract}

\newpage

\section {Introduction}

Quantum computers are hypothetical devices
based on quantum physics. A formal definition of
quantum computers 
was pioneered by
Deutsch \cite {D}, who also realized that they can outperform
classical computation.\footnote{  The idea of a
quantum computer can be traced back to works by
Feynman, Manin, and others, and this development is also related to
reversible computation and connections between computation and physics 
that were studied by Bennett in the 1970s.}
Perhaps the most important result in this field and
certainly a major turning point was
Shor's discovery \cite {S1} of a polynomial quantum
algorithm for factorization. The notion of a quantum computer 
along with the associated complexity class BQP
has generated a large body of research in theoretical and experimental 
physics, computer science, and mathematics.  
For background on quantum computing, see
Nielsen and Chuang's 
book \cite {NC}.

Of course, a major question is whether quantum computers are
feasible. An early critique of quantum computation (put forward in
the mid-90s by 
Landauer \cite {lan,lan2}, Unruh \cite {unr}, and others) 
concerned the matter of noise:

\begin {itemize}
\item [{\bf [N]}]
{\bf The postulate of noise: Quantum systems are noisy.}
\end {itemize}

The foundations of noisy quantum computational 
complexity were laid by Bernstein and Vazirani in \cite {BV}.
A major step in showing that noise can be handled was the
discovery by Shor \cite {S2} and Steane \cite {St} 
of quantum error-correcting codes.
The hypothesis of fault-tolerant quantum computation (FTQC) was
supported in the mid-90s by the ``threshold theorem'' proved by 
Aharonov and Ben-Or \cite {AB2}, Kitaev \cite {K1}, 
Knill, Lafflame, and Zurek \cite {KLZ},
and Gottesman  \cite {Got}. The threshold theorem asserts
that under certain natural assumptions of statistical 
independence on the noise, if the rate of noise (the amount of noise
per qubit in one computer cycle) is a small constant, 
then FTQC is possible. It
was also proved that high-rate noise is an obstruction to FTQC.

The study of quantum error-correction and its limitations, as well as 
of various approaches to fault-tolerant quantum
computation, is extensive and beautiful; see, e.g., \cite
{AB1,Got,ABIN,CS,K2,Kn,Ra}.


Concerns about noise models with statistical dependence are 
mentioned in several places, e.g.,
\cite {Pr,Le}, and specific models of noise that may be 
problematic for quantum error-correction were studied by Alicki, 
Horodecki, Horodecki, and Horodecki \cite {AHHH}.
Current FTQC methods apply to even more general models of noise
than those first considered, which allow various forms of time-
and space-statistical dependence; see 
\cite {TB,AGP,AKP}.

The purpose of 
this paper is to present two conjectures
concerning decoherence for quantum computers  
which, if (or when) true, are damaging for quantum error-correction 
and fault-tolerance.

The first conjecture concerns entangled 
pairs of qubits.

\begin {itemize}
\item [{\bf [A]}]
A noisy quantum computer is subject to error with the property that 
information leaks for two substantially
entangled qubits have a substantial positive 
correlation.
\end {itemize}

We emphasize that Conjecture [A] 
refers to part of the overall error affecting a noisy quantum computer. We refer to this error as 
{\it detrimental}. Other 
forms of errors and, in particular, errors consistent
with current noise models may also be present. (We conjecture that the effects
of detrimental errors described by Conjectures [B] and [C] below cannot 
be remedied by additional errors of a different nature.)

Error synchronization refers to a situation where,
while the error rate is small, there is a substantial probability of errors
affecting a large fraction of qubits.

\begin {itemize}
\item [{\bf [B]}]
In any noisy quantum computer at a highly entangled state there will be a
strong effect of 
error synchronization.
\end {itemize}

We will now describe the structure of the paper. 
Section \ref {s:n} gives more background on noise and fault-tolerance.  
In Section \ref {s:e} we define 
a simple class of errors that suffice to demonstrate 
our conjectures and to show the connection between Conjectures [A] and [B].

The main Section \ref {s:mf} is devoted to mathematical formulations 
of the above conjectures and we also discuss why these conjectures are
damaging to quantum error-correction.  In the Appendix, stronger versions of 
Conjecture [A] are formulated. 

Section \ref {s:mod} discusses 
examples that may give 
the conjectured behavior and 
actual models of noise 
that may demonstrate this behavior. 
We first discuss
error propagation 
of ``unprotected 
quantum programs'' as a prototype for detrimental errors. A main thesis 
of this paper is that ``new errors'' in the evolution of a quantum computer 
are modeled after the behavior of unprotected quantum circuits. 

Section \ref {s:cs} discusses 
related aspects of computational complexity. While our conjectures
appear to be damaging to current fault-tolerance methods based on quantum 
error correction, it is still plausible that 
under the assumption of low-rate errors 
conjectures [A] and [B] will allow log-depth quantum computation. 

Section \ref {s:ext} discusses extensions 
of these conjectures to more general quantum systems and briefly touches on 
relations with classical noisy systems. Section \ref {s:rate} discusses 
the rate of noise. Section \ref {s:con} concludes.

The conjectures of this paper can be regarded as proposed 
properties for error models for quantum computers (and, at a later stage, 
for more general quantum systems) that will cause 
quantum error-correction and FTQC to fail.  Alternatively, the 
conjectures can be regarded as proposed {\it consequences} of 
lack of fault-tolerance in quantum 
systems. As such they can be relevant 
to the nature of decoherence of 
quantum physical systems in nature even if quantum computers are possible.
At present, there are no clear examples 
of quantum error-correction or of quantum 
fault-tolerance in quantum processes in nature.

We list now (again in an informal manner) additional conjectures 
made in this paper.


1. (Section \ref {s:mf}): We will refer to a pure state of a quantum computer 
that up to a small error is induced
by its ``marginal distribution'' on small sets of qubits as
``approximately local.'' (See Section \ref {s:mf} for a formal definition.)

\begin {itemize}
\item [{\bf [C]}]
The states of noisy quantum computers are approximately local.
\end {itemize}

2. The next conjecture proposes an extension of the 
conjectures above to general quantum systems (Section \ref {s:ext}).
\begin {itemize}
\item [{\bf [D]}] 
A description (or prescription) of a noisy quantum system 
at a state $\rho$ is subject to error given by 
a quantum operation $E$ that
tends to commute with every
unitary operator that stabilizes $\rho$.
\end {itemize} 

(Here, ``tends to commute'' reflects a small bias towards 
commutativity which will
be motivated and described further in Section \ref {s:ext}.) 

3. (Section \ref {s:rate}) 
The following further conjecture has some bearing also on the rate 
of noise and can be regarded as 
a strong form of the postulate of noise itself. 

\begin {itemize}

\item [{\bf [E.1]}]
A noisy quantum system
is subject to (detrimental) noise with the following property: 
the infinitesimal rate of noise at time $t$ (in terms of trace distance) 
is 
bounded from below by a measure of non commutativity between 
the operators describing the
evolution 
prior to time $t$ and those describing it after time $t$.  

\item [{\bf [E.2]}]
A noisy quantum computer with $n$ (logical) qubits 
is subject to (detrimental) noise, with the following property: 
For some $\kappa>0$, 
the infinitesimal rate of noise 
(in terms of trace distance)  
is at least  
$\kappa n.$

\end {itemize}

For standard models of noise, the infinitesimal rate of noise 
in terms of trace distance scales up linearly with the number of qubits. 
However, in cases of highly entangled
states, which, by Conjecture [B], would lead to error-synchronization,  
conjecture [E.2] suggests that the rate of detrimental 
noise for individual qubits 
will scale up linearly with the number of involved qubits. 
Conjecture 
[E.2] 
runs counter to the possibility of ``decoherence 
free subspaces,'' 
which are possible for the standard noise models.



\section {Quantum computers, noise, and fault-tolerance}
\label {s:n}

The state of a digital
computer having $n$ bits is a string of length $n$ of zeros and ones.
As a first step towards quantum computers we can consider 
(abstractly) stochastic versions of digital computers where the state is 
a (classical) probability distribution on all such strings. 
Quantum computers are similar to these (hypothetical) stochastic 
classical computers and they work on qubits (say $n$ of them). The state of 
a single qubit $q$ is described by a unit vector $u = a|0>+b|1>$  in 
a two-dimensional complex space $U_q$. 
(The symbols $|0>$ and $|1>$ can be thought of as 
representing two elements of a basis in $U_q$.) 
We can think of the qubit $q$ as representing 
$`0'$ with probability $|a|^2$ and $`1'$ with probability $|b|^2$. 
The state of the entire computer is a unit vector in the $2^n$-dimensional 
tensor product of these vector spaces $U_q$'s for the 
individual qubits. The state of the computer thus 
represents a probability distribution on the $2^n$ strings
of length $n$ of zeros and ones. The evolution of the quantum 
computer is via ``gates.'' Each gate $g$ 
operates  
on $k$ qubits, and we can assume $k \le 2$. 
Every such gate represents a 
unitary operator on $U_g$, 
the ($2^k$-dimensional) tensor product 
of the spaces that correspond to these $k$ qubits. At every ``cycle time''
a large number of gates acting on disjoint sets of qubits operate.

Moving from a qubit $q$ to the probability distribution on $`0'$ and $`1'$
that it represents is called a ``measurement'' and
it can be considered as an additional 1-qubit gate. We will assume
that measurements of qubits that amount to a sampling of 0-1 strings 
according to the distribution these qubits represent is the final step 
of the computation.


The postulate of noise is essentially a hypothesis about
approximations. The state of a quantum computer can be
prescribed only up to a certain error. For FTQC there
is an important additional assumption on the noise, namely, on the
nature of this approximation.  The assumption is that the noise is
``local.'' This condition asserts that the way in which the 
state of the computer changes between 
computer steps is statistically independent, for different qubits. 
We will refer to such changes as ``storage errors'' or as ``qubit errors.'' 
In addition, the gates that carry the computation itself are imperfect.
We can suppose that every such gate involves a small number 
of qubits and that the gate's
imperfection 
can take an arbitrary form, 
and hence the errors (referred to as ``gate errors'') created on the few qubits 
involved in a gate can be statistically dependent. 
Of course, qubit errors and gate errors propagate along the computation. 

The basic picture we have of a noisy computer is that 
at any time during the computation 
we can approximate 
the state of each qubit only up to some small error term 
$\epsilon$. Nevertheless, under the assumptions concerning the errors 
mentioned above, computation is possible. The noisy physical qubits
allow the introduction of logical ``protected'' qubits that are 
essentially noiseless.

Our conjectures apply to the same model of quantum computers 
but they require a more general notion of errors. They 
require that the storage errors 
will not be statistically independent (in fact, they should 
instead be very dependent) 
or that the gate errors will not be 
restricted to the qubits involved in the gates and will 
be of sufficiently general form. 
(Note that the errors
may also reflect the translation from this ideal notion of quantum 
computers to a physical realization.) 

\section {A simple form of error}
\label {s:e}

\subsection {Simple forms of error}

Let $W_k$ be the error operation 
that amounts to changing the state of the $k$th 
qubit to $\rho_0$, the maximum entropy state.  
Let $I_k$ be the identity operation for the $k$th qubit. 
For a 0-1 vector $x=(x_1,x_2,\dots,x_n)$  let 
$ E_x = \otimes_{k=1}^n E_k$ where $E_k=W_k$ if $x_k=1$, 
and $E_k=I_k$ if $x_k=0$.
 
Let $\cal D$ be a probability distribution on 0-1 vectors
of length $n$. We let $E_{\cal D}= \sum {\cal D}(x)E_x$ where 
the sum is taken over all 0-1 vectors $x$ of length $n$. We will refer 
to errors of the form $E_{\cal D}$ as {\it simple}. 
 
Simple errors are useful in demonstrating
the notions we discuss. The standard assumption regarding 
storage noise would mean, in this special case, that ${\cal D}$ is a 
product probability distribution.
 
A case of particular interest is when the probability distribution 
${\cal D}(x)$ depends only on the number of `1's in the vector $x$.
Another way to describe this special case is to consider the error 
(regardless of the number $n$ of qubits) as depending on a 
probability distribution $f$ on $[0,1]$. We first choose a real number $t$,
$0 \le t \le 1$, according to the distribution $f$ and then, for every $k$, 
change the state of the $k$th qubit to $\rho_0$, with probability $t$ 
(independently for different qubits). 

\subsection {The amount of error, error synchronization, and correlation}

A measure for the amount of error (or information leak) 
for the $k$th qubit for the error operation
$E_{\cal D}$ is just $p_k({\cal D})$, defined as the probability that $x_k=1$. 
For the special case $E_f$ the amount of error for every qubit is 
$R(f)=\int_0^1f(t)dt$.
    
It is simple to describe the notion of error synchronization in this setting.
For errors of the form $E_f$, error synchronization 
refers to a situation where $R(f)$ is small, but for some 
$t>> R(f)$, $\int _t^1 f(t)dt$ is substantial. 
When the error is described by $E_{\cal D}$, error synchronization refers 
to the situation where $p_k({\cal D}) \le t$ for every $k$ and a 
small real number $t$, but the probability for $x=(x_1,x_2,\dots,x_n) \in {\cal D}$, with
$x_1+x_2+\cdots+x_n> sn$, is substantial for $s>>t$.




For a probability distribution $\cal D$ and two qubits $j$ and $k$ 
let $c_{jk}({\cal D})$ be the correlation between the events
$x_j=1$ and $x_k=1$. When we consider a block of qubits 
representing an error correction code, then (under some 
additional assumptions see; Section \ref {s:mf}) 
Conjecture [A] asserts that these pairwise 
correlations are high. For error operations 
of the form $E_{\cal D}$ it is easy to deduce 
error synchronization from high pairwise correlation:

\begin {lemma}
\label {l:LEMMA}
Let $t<1/20$ and $s>4t$.
Suppose that $\cal D$ is a distribution of 0-1 strings of length $n$
such that the $p_i({\cal D}) \ge t$ and $c_{ij}({\cal D}) \ge s$. Then
\begin {equation}
{\bf Prob} (\sum_{i=1}^n x_i > sn/2) > st/4. 
\end {equation}
\end {lemma}

{\bf Proof:} It is easy to see that the probability in question is minimized
if ${\cal D}$ is symmetric with respect to permutations of the variables. 
Suppose that for such a symmetric distribution 
we want to minimize the pairwise 
correlation subject to the condition that 
the probability that $x_1+x_2,+ \cdots +x_n > s/2$ is at 
most $st/4$, and the probability that every $x_i=1$ is at most $t$. 
It is easy to see that we should choose the following distribution: 
with probability $p_1=ts/4$ all $x_i's$ are 1. With probability 
$p_2$ 
we do the following: we choose uniformly at random a 
set $R$ of size $[sn/2]$ and let $x_i=1$ if $i$ belongs to this set. 
We choose $p_1$ and $p_2$ so that the probability for $x_i=1$ is $t$. 
Finally, with 
probability $1-p_1-p_2$ we choose all $x_i=0$. Now, slightly 
modify this distribution as follows: 
with probability $st/4$, choose $x_i=1$ for every $i$; 
with probability $1-2t/s+t/2-st/4$, choose $x_i=0$ for every $i$; and 
with probability $2t/s-t/2$, choose, (independently for different $i$s,) 
$x_i=1$ with probability $s/2$.

The probability of each individual $x_i$ being 1 is still $t$ and 
the pairwise correlations did not decrease. And now we can easily 
compute the pairwise correlation and they turn out to be smaller 
than $s$. (As pointed out by Yuval Peres, this type of 
lemma falls under the known analysis 
of the Curie-Weiss model \cite {Ellis}.
  



\section {A mathematical formulation} 
\label {s:mf}


In this section we give mathematical 
formulations for Conjectures [A], [B], and [C]. 
Our setting is as follows. We have a quantum computer running on $n$ qubits.
The ideal (or ``intended'') state of the computer is pure. 
We want to propose a picture for noisy quantum computation based 
on this basic model of a (noiseless) quantum computer. We assume 
that the actual state of the computer is close to the ideal state $\rho$. Our 
conjectures refer to the ``new errors'' (storage and gate errors) in 
one computer cycle.


The error can be described
by a unitary operator on the computer qubits and the neighborhood qubits
or by a quantum operation $E$ on the space of 
density matrices for these $n$ qubits.  
We will not give 
a specific model of detrimental error but rather 
describe some of its expected properties.

\subsection {Two qubits}

We first describe a measure of information leak. 
For a state $\rho$ of the computer and a set $A$ of qubits
let $\rho|_A$ be the induced state on $A$. 

Consider a quantum operation $E$.
Note that when the state $\tau$ of the quantum computer is a tensor 
product pure state  
then for every set $A$
of qubits, $S(\tau|_A)=0$. Here, $S(*)$ is the (von Neumann) entropy function; 
see, e.g., \cite {NC}, Ch. 11.
The information leak of 
the noise operator $E$
from the set of qubits $A$, w.r.t. $\tau$, can be measured by 
the entropy $S((E(\tau)|_A)$.  
For a tensor product state $\tau$ and a qubit $a$ define 
$L_E(a;\tau) = S(E(\tau)|_a))$; more generally, for a set $A$ of 
qubits define
$$L_E (A;\tau) = S(E(\tau)|_A)).$$



We will now state mathematically a version of Conjecture [A].
Our setting is as follows. Let $\rho $ be the ``intended'' (``ideal'') pure
state of the computer and consider two qubits $a$ and $b$. 
We  use as the (rather standard) measure of entanglement 
between qubits at pure states

$$ ENT(\rho; a,b)= S(\rho |_a)+S(\rho |_b) - S(\rho |_{ \{a,b\} }).$$
\noindent
As a measure of correlation of information leaks we use
$$EL_E(a,b;\tau) = L_E(a;\tau)+L_E(b;\tau)-L_E(\{a,b\};\tau).$$

Conjecture [A] can be formulated as follows: 

For every tensor product state $\tau$, 

\begin {equation}
\label {e:p1}
EL_E(a,b;\tau) \ge  K(L_E(a;\tau),L_E(b;\tau)) \cdot ENT(\rho; a,b),
\end {equation}

\noindent
where $K(x,y)/\min (x,y)^2 > > 0$  
when $x$ and $y$ are 
positive and small.
($K (x,y)=0$, when $\min (x,y) = 0$ and so  
relation (\ref {e:p1}) tells us nothing about 
noiseless entangled qubits.) 


Below in Section \ref {s:ee} we will describe and motivate 
a stronger form of Conjecture [A] based on a different measure 
for entanglement. In the appendix we point out 
an alternative mathematical formulation for information leaks 
and describe an extension of Conjecture [A] to several qubits rather than two.

A simple extension that we would like to mention at this point
is to pairs of {\it qudits}
rather than pairs of qubits. The term qudit is used to denote a unit 
of quantum information in a $d$-level quantum system. 
Relation (\ref {e:p1}) extends to 
qudits without any change.  
This applies, in particular, to two disjoint sets of qubits 
in a quantum computer. 

{\bf Remark:} Consider two qudits $a$ and $b$, 
with $d$ and $d'$ possible levels respectively. 
The ideal pure state of this  pair of qudits is represented by 
a $d$ by $d'$ matrix. 
Our conjecture (roughly) asserts that when the state is not represented by 
(or close to) a rank one matrix then neither is the error.  






\subsection {Error synchronization}

A simple way to describe error synchronization is in terms of 
the expansion of the quantum operation E in terms of multi-Pauli operators.
A quantum operation $E$ can be expressed as a linear combination 
$$E=\sum v^I P^I,$$ 
where $I$ is a multi-index $i_1,i_2,\dots,i_n$, 
where $i_k \in \{0,1,2,3\}$ for 
every $k$,  $v^I$ is a vector, and $P^I$ is the quantum 
operation that corresponds to 
the tensor product of Pauli operators whose action 
on the individual qubits is described by the multi-index 
$I$. The amount of error on the $k$th qubit is described by
$\sum \{ \|v^I\|_2^2: i_k \ne 0 \}$. For a multi-index $I$ 
define $|I|=|\{k:i_k \ne 0\}|$. Let 
$$f(t) =: \sum \{ \|v^I\|_2^2: |I| = t\},$$
we can regard $\int_{0}^1 f(t) t$, the average over the qubits of the 
amount of error, as the error rate.

Suppose that the 
error rate is $a$. 
All noise models studied in the 
original papers of the ``threshold theorem,'' 
as well as some extensions that allow time- and 
space-dependencies (e.g., \cite {AKP}), have the property that 
$f(t)$ decays exponentially (with $n$) for  $t=(a+\epsilon)n$, 
where 
$\epsilon >0$ is any fixed real number.  

In contrast, we say that $E$ leads to {\it error synchronization} if $f(\ge t)$ is substantial 
for some $t >> a$. We say that 
$E$ leads to a {\it strong} error synchronization if 
$f(\ge t)$ is substantial for 
$t=1/2-\delta$ where $\delta = o(1)$ as $n$ tends to infinity, 
and to {\it very strong} 
error synchronization if  $f(\ge t)$ is substantial for 
$t=3/4-\delta$ where $\delta = o(1)$ as $n$ tends to infinity.
A random unitary operator on the qubits of the computer with or without 
additional qubits representing the environment yields very strong 
error synchronization.



\subsection {Conjectures [A] and [B], quantum error-correction, 
and fault-tolerance}
\label {s:abqec}

For a fault-tolerant quantum computer (for current methods 
of fault-tolerance), when the 
number of qubits is large, for most pairs of qubits the errors at 
every time of the computation, namely the gap between the intended 
state and the actual state,  will be almost statistically independent.   
This property is assumed for the ``new errors'' (the storage 
errors and the gate errors combined), both for the standard model of noise and 
for recent, more general models of noise \cite {AKP}. Remarkably, 
when this property is satisfied for the new errors, and the error rate 
is small, fault-tolerant schemes allow us to keep this property 
for the accumulated error.

I will now describe why 
Conjecture [A] (or rather an appropriate strengthening) is  
damaging 
for quantum error-correction and FTQC. 
We will first consider two simplifying assumptions:
\begin {enumerate}

\item
Measuring a qubit and looking at its content does not induce 
errors on other qubits.

\item 

The error 
is of the simple form $E_{\cal D}$.

\end {enumerate}

Consider the state of a quantum computer that applies 
a fault-tolerant computation.
The state of the computer (or of large blocks of qubits of the computer) 
is $t$-wise independent for a large value 
of $t$; hence every two qubits are statistically independent and 
Conjecture [A] does not directly apply.
Consider an error-correcting code and let $s$ be the minimal number 
of qubits whose state ``determines'' that of the others, so that once 
they are measured and their values are ``looked at'' the state of the 
other qubits is determined. When we measure and look at the values 
of $s-1$ qubits, we see a very strong dependence between 
every pair of  the remaining qubits. Given our first assumption, 
measuring the other qubits does not affect the error on 
the two qubits we are interested in. Therefore, Conjecture [A] implies that 
the correlation of 
information leaks for every pair of qubits is substantial. If the error 
has the simple form $ E_{\cal D}$ then Lemma \ref {l:LEMMA} asserts 
that there is a strong form of error synchronization and this 
will fail the quantum error=correction required for the 
threshold theorem. 


In order to extend the above argument so that it will not be 
based on assumption (1) we propose in Section \ref {s:ee} below 
a stronger form of Conjecture [A] that 
relies on a notion of ``emergent entanglement.'' 

Assumption (2) poses a serious 
limitation to the above argument,
but I expect that reliance on assumption (2) is technical and  
that Lemma \ref {l:LEMMA} extends to general forms of errors.

We now show that a certain simple error model that satisfies 
Conjecture [A] (and even the strong conjectures of the Appendix) 
and Conjecture [B] (and even the strongest version of error-synchronization) 
still allows the use of log-depth quantum 
computation, e.g., for polynomial time factoring. 

The model is very simple. In each computer cycle 
with probability $1-t$ nothing happens 
and with probability $t$ every qubit collapses to its 
maximum entropy mixed state.
(In other words, the new errors are described by $E_{\cal D}$, where $\cal D$ 
is the distribution giving the all 0 vector probability $1-t$, and 
the all 1 vector probability $t$.)

If we run a log-depth quantum circuit a polynomial number 
of times, one of the runs will work without any error. For an algorithm 
like factoring, if we run the algorithm 
including the quantum subroutine a polynomial number 
of times, at one of these times we will end up 
with a correct factoring that we will be able to check  
in polynomial time.

{\bf Remark:} Following is a simple 
argument proposed by Kuperberg why 
even the simplest form of Conjecture [A] 
would not allow quantum computation at all.
``If quantum computing is possible, then a quantum computer
could have prepared a state $S$ and then communicated it to the system that
has the noise operation $E$.  If it is true quantum computing, then $S$
can be secret from $E$, for reasons similar to those that make quantum key
distribution possible.  In this case $E$ can act on $S$ but it cannot
otherwise depend on it.''
The difficulty with this argument 
is that moving from a logical 
protected state $S$ to a physical realization of $S$ on a different device
requires some computation and fault-tolerance and thus relies on assumptions
regarding errors that we cannot assume. Still, Kuperberg's
proposed reduction can be useful. (The example above 
shows that we cannot expect such a strong statement as Kuperberg's.)

\subsection {Two qubits: emergent entanglement}

\label {s:ee}

We proceed to describe and motivate a stronger form of 
Conjecture [A]. 

The expression $S(\rho|_a)+S(\rho|_b) - S(\rho|_{\{a,b\}})$ was used 
as a measure of entanglement between two qubits. We would like 
to replace it by a measure that can be 
called ``emergent entanglement,'' which we are now going to define.
This measure, denoted by $EE(\rho; a,b)$, captures (roughly)  
the expected amount of entanglement among the two qubits 
when we measure some other qubits, ``look at the outcome,'' and condition
on all possible outcomes for the measurement.
It appears to be related to Briegel and Raussendorf's 
notion of ``persistent entanglement'' \cite {BR}. 




For every representation $\omega$ of $\rho|_{ \{ a,b \} }$
as a mixture (convex combination) of 
joint states 
$$\rho|_{ \{ a,b \} } = \sum_{i=1}^t p_k \rho _k,$$ let 
$$ENT_\omega (\rho ;a,b ) = \sum p_k ENT(\rho_k;a,b).$$  
Define $$EE(\rho; a,b) = \max ENT_\omega (\rho; a,b), $$ where the maximum
is taken over all representations $\omega$. (We can assume that $\omega$ 
is a mixture of pure joint states.)

A strong form of relation (\ref {e:p1}) is
\begin {equation}
\label {e:p1strong}
EL(a,b) \ge  K(L(a),L(b)) \cdot EE(\rho; a,b),
\end {equation}

\noindent
where, as before, $K(x,y)/\min (x,y)^2 > > 0$  
when $x$ and $y$ are 
positive and small. 






Using the measure for emergent entanglement appears to give 
the ``right'' formulation for Conjecture [A]. It may also be relevant  
for formulating 
our conjectures when the intended state is not necessarily pure. 
Aharonov \cite {A:new}
proved that quantum computers with mixed states can run arbitrary 
quantum computation without any entanglement between qubits of the computer.
(However, these qubits will have high emergent entanglement.)


\subsection {Censorship}

Here is a suggestion for an entropy-based 
mathematical formulation for Conjecture [C]. We remind the reader that 
in this section  we always assume that the ``ideal'' state of the quantum 
computer (before the noise is applied)
is a pure state. Some adjustments to our conjectures 
will be required when the 
ideal state itself is a mixed state. 

Let $\rho$ be a pure state on a set 
$A = \{ a_1,a_2,\dots,a_n\} $ of $n$ qubits. Define

$$ENT(\rho; A) = -S(\rho)+ \max S(\rho^*),$$ 
where $\rho^*$ is a mixed state with the same marginals
on proper sets of qubits as $\rho$, i.e., 
$\rho^*|_B = \rho|_B$ for every proper subset $B$ of $A$.

Next, define

$$\widetilde{ENT}(\rho) = \sum \{ENT (\rho; B): B \subset A \}.$$

\noindent
In this language a way to formulate the censorship conjecture is:

\medskip

\noindent
Conjecture {\bf [C]}: 
There is a 
polynomial $P$ (perhaps even a quadratic polynomial) such that 
for any quantum computer on $n$ qubits, 
which describes 
a pure state $\rho$, 
\begin {equation}
\label {e:c}
\widetilde{ENT} (\rho) \le P(n). 
\end {equation}

\medskip


\section {Examples and models}
\label {s:mod}
\subsection {Unprotected quantum circuits}
\label{s:un}

A basic example to have in mind (with some caveats that we 
will discuss below) is the example
of ``unprotected'' quantum circuits. This goes back 
to Unruh \cite {unr} who 
studied error propagation for Shor's algorithm. 
Take the standard model of 
statistically independent errors and 
suppose that the error rate is so small  that it accumulates 
at the end of the computation to a small constant-rate error.  
It is instructive to see 
in this context 
that error synchronization 
is often created. 

We emphasize that 
since fault-tolerant quantum computation handles well propagation of errors,
a model for decoherence that supports Conjectures [A] and [B] (and [E.2], 
Section \ref {s:rate} )
should already exhibit 
[A] and [B] (and [E.2]) 
for the ``new errors'' --- whether storage errors 
or gate errors. 






A main (yet informal) 
thesis of this paper is that a noisy quantum computer   
is subject to a substantial amount of error  
that behaves like 
propagated error for unprotected quantum circuits.

There are no (definite) examples of quantum error-correction in 
nature, and this suggests that
models of unprotected quantum circuits can suffice to 
represent the 
evolution of natural quantum processes. Therefore, a study of the evolution 
of unprotected quantum circuits is of 
independent interest.

Let me remark at this point that errors of unprotected quantum 
circuits leading to a 
state $\rho$ will exhibit systematic dependencies on $\rho$ of 
various forms. 
When we have a discrete set of gates we encounter errors that depend 
discontinuously on $\rho$. This is an issue that has been studied  
in the quantum information literature and is related to 
the Kitaev-Solovay Theorem, see \cite {NC}; Chapter 4 and Appendix 3. 
In this paper, we 
are interested in a different phenomenon, that of 
systematic smooth dependence of the errors on $\rho$. 
(This justified our description of the errors in 
terms of quantum operations to start with.) 
For this purpose it seems reasonable to consider quantum circuits with 
a continuous set of gates. 

We should offer a 
more precise definition of ``unprotected quantum circuits.'' A random circuit 
leading to a given state $\rho$ or a random perturbation of a 
specific circuit leading to $\rho$ (which still leads to $\rho$) may 
serve this purpose.

Next we should describe the model of noise. 
To start, we may consider the standard model of independent 
noise (even the very simple model, considered in Section \ref {s:e},  
when every qubit collapses to a fixed state 
with a small probability, and these probabilities 
are statistical independent), 
and study the accumulation of errors when the noise for a computer cycle 
is very small.\footnote {However, note that we 
conjecture that new errors in a process 
leading to a state $\rho$ share properties with accumulative errors of 
an unprotected program leading to $\rho$ and, therefore, taking the 
standard model of statistically 
independent errors to understand unprotected 
circuits can be regarded as a first approximation.}

\subsection {Models}
\label {s:mod-s}

A  basic remaining challenge is to present concrete models of noise
that support our conjectures.

We Emphasize again that 
a model for decoherence that supports conjectures [A] and [B] 
should already exhibit 
[A] and [B] 
for the ``new errors'' --- whether storage errors 
or gate errors\footnote {As mentioned, 
we should allow gate errors to ``apply'' also 
to qubits not involved in 
the gate. Allowing this may reflect several concerns 
expressed in the literature regarding the qubit/gate model such as the issue 
of ``slow'' gates \cite {ALZ}.}
or both ---
and thus be quite different from 
current models and current perceptions regarding noise. 
Models that satisfy our conjectures
may be based on the storage errors (in a single computer cycle) 
being represented  by a rather
primitive (but quick) stochastic quantum program (or circuit). Here are 
a few additional points regarding concrete noise models that may be relevant. 

1) Noise models satisfying our conjectures can be regarded as 
a further step in the direction considered recently by 
Aharonov, Kitaev, and Preskill \cite {AKP} (and a few earlier works).
In these works, interactions between nearby qubits (arranged on a grid) 
that lead to statistical dependence between the noise acting on 
them are considered and it is shown that the threshold 
theorem prevails if the independence assumption still applies to faraway 
qubits. 
However, 
interactions between nearby qubits 
may lead to dependencies between 
errors that are not covered by the assumptions of \cite {AKP}.
(Compare the remark about cluster states below.)


2) Klesse and Frank \cite {KF} described 
a physical system in which 
qubits (spins) are coupled to a bath of massless bosons and 
then reached (after certain simplifications) a noise model with 
error synchronization.  

3)  The earlier models suggested by Alicki, Horodecki, Horodecki, and Horodecki
\cite {AHHH} appear to be relevant to our conjectures. 

4) Let me also mention the relevance of {\it cluster states} defined by 
Briegel and Raussendorf (see \cite {RBB}).  
The description of cluster states involves 
an array of qubits located on the vertices of a rectangular
lattice in the plane (or in space).
Cluster states are ``generated'' by 
local entanglement between pairs of nearby 
qubits on the lattice grid. 
They can be regarded as the quantum analogs of the Ising 
and Potts classical models. 

Controlled creation and manipulation of cluster states can be
important for building quantum computers. 
On the other hand, 
cluster states and the local processes leading to them
can possibly serve as a basis
for concrete models of detrimental decoherence. 


5) A toy model for noise that neglects some of the 
effects we consider in the paper 
and brings others to an extreme form is the following.
There are no gate errors. Consider the graph $G$ 
whose vertices are the qubits and whose edges are qubits 
that occur in a gate. Edges are labeled by the gate imperfection. 
The storage error is described by $E_{\cal D}$ where the probability 
distribution $\cal D$ is  given by an Ising model on the graph $G$ 
based on these gate imperfections.  
Can quantum error-correction prevail in such (low-rate) error model? 

\section {Computation complexity} 
\label {s:cs}

Scott Aaronson's interesting  ``Sure/Shor challenge'' \cite {Aa1} 
asks for restrictions 
on feasible (physical) states for 
quantum computers that do not allow for polynomial time factoring of integers
and at the same time do not violate what can already be 
demonstrated empirically.  
This looks like a difficult challenge. 
In a similar spirit, while it looks intuitively correct that 
our conjectures are
damaging to quantum computation, proving it  
is not going to be easy.

A realistic task would be to show 
that our conjectures exclude 
fault tolerance based on linear quantum 
error-correction, 
e.g., deriving relations 
(\ref {e:p1}) and (\ref{e:c}) (or even (\ref{e:t2c})) 
for any form of ``protected qubits'' obtained by
linear quantum error-correction.

A more ambitious goal than excluding quantum linear error-correction
would be finding a reduction of noisy quantum computation (with 
detrimental errors) 
to the computational power of log-depth quantum circuits. 
(This will still fall short of Aaronson's challenge
in view of Cleve and Watrous \cite {CW}, who gave a polynomial  
algorithm for factoring that requires, beyond classical computation, only 
log-depth quantum computation.) 
Reductions to log-depth quantum computation 
are known under the standard assumptions on noise, for reversible 
quantum computation \cite {ABIN}. For certain noise models, 
when the error rate is 
above 45\% \cite {BCLLSU}, it is known that the noisy quantum 
computer can be simulated by a classical computer. 

When we insist on small error rate it may well be the case that 
log-depth 
quantum circuits represent the true complexity power of quantum 
computers with detrimental errors. As we already pointed out 
in Section \ref {s:abqec}, for a log-depth circuit such that 
the storage (and gate) errors
demonstrate perfect error synchronization
running the computation a polynomial number of times, with high probability
there will be no errors in one of the runs. 
If we replace a given log-depth circuit by a larger one capable of 
correcting standard errors we may reach
polynomial size (or quasi-polynomial size) circuits that are immune to 
low-rate errors of the kind considered in this paper.
(Conjecture [E.2] below 
may be damaging also to log-depth quantum computation.)



\section {Extensions}
\label {s:ext}
\subsection {General quantum systems}

The purpose of Section \ref {s:mf} was to describe formally 
the conjectures on decoherence of quantum computers based on the 
basic model of such a computer. In the context of 
general quantum systems these conjectures are 
thus somewhat arbitrary. (In particular, we always talk 
about Hilbert spaces of dimensions $2^m$.)
  
The main idea behind the conjectures is that the 
error-independence assumption (for different qubits) 
amounts to an extremely strong 
dependence of the errors on the tensor product
structure of the Hilbert space describing the state of the computer. 
It can  
be useful to suggest and examine 
formulations of our conjectures that do not depend on the tensor product
structure of the Hilbert space in question.


We want to consider quantum physical systems described by a complex 
Hilbert space $V$.
Our conjectures suggest that
if $E$ represents the error for state $\rho$ and $E'$ represents 
the error for state $U(\rho)$, 
for a unitary operator $U$ on $V$, then $E'$  
will be ``close'' to $U^{-1}EU$. In particular, this implies that 
if $U(\rho)=\rho$ then $E'$ is ``close'' to $U^{-1}EU$; 
hence $UE$ is ``close'' to $EU$.
In other words, $E$ and $U$ ``tend'' 
to commute if $U(\rho)=\rho$.

Here is an attempt at a formal conjecture.
We will restrict our attention to the special case where the error 
is described by a quantum operation $E$  
which is a convex combination of unitary operators.

\begin {itemize}
\item [{\bf [D]}] 

There is an $\alpha >0$ such that 
a prescription (or description) 
of a noisy quantum system 
at a state $\rho$ is subject to error 
$E$ having the property that for every 
unitary operator $U$ such that $U(\rho ) = \rho$ 

\end {itemize}

\begin {equation}
\label {e:t2c}
\| EU-UE \| \le (1-\alpha) \sqrt 2.
\end {equation}  

Here we do not insist that the prescribed (or described) 
state be pure, and we refer to the Hilbert-Schmidt norm.

Several remarks are in place. First, 
Greg Kuperberg pointed out that at a thermodynamics equilibrium 
a certain limiting error $E$ will actually commute with every $U$ 
that stabilizes $\rho$. We can regard Conjecture [D] 
as a statement referring to non equilibrium 
thermodynamics.\footnote{In this context, the works (and even the 
small controversy) on quantum analogs of ``Onsager's regression theorem'' 
come to mind.} 

Second, while a generic form of noise (say, a generic unitary 
operator with prescribed rate) indeed leads to error synchronization and
is damaging to quantum error-correction, such a noise appears unrealistic. 
Here, the condition on the noise is nongeneric, and rather 
it is the standard assumption 
on noise that leads (for highly entangled states) 
to generic commutativity behavior between the 
noise operation and the state of the computer.\footnote{Going back to 
the issue of approximating arbitrary matrices up to rank one matrices, 
note that if $A$ is a random 
$n$ by $n$ matrix and $D$ is {\it any} rank one matrix then 
with high probability
$\|AD-DA\| \ge (1-o(1))\sqrt 2$.}   

Third, it will be interesting to examine 
Conjecture [D] for noisy adiabatic models of quantum computers; see, 
e.g. \cite {CFP}. 




\subsection {Classical noise}

Conjectures [A] and [B] were 
originally formulated in 
\cite {Ka2}
also for  ``natural'' noisy classical 
correlated systems. 
For example, the analog of [A] asserts that 
in a noisy system the errors for two highly correlated elements tend to be
substantially correlated.
Because of the heuristic (or subjective) nature of the notion 
of noise in classical systems (and of the notion of probability itself),
such a formulation, while of interest, leads to several difficulties.

Understanding noise and the study of de-noising methods span wide areas. 
(For example, in machine learning we can see the example 
where text and speech represent respectively the intended (ideal) and 
noisy signals.) 
Certain statistical 
methods of de-noising
are based on assumptions that run counter to [A]. However, our conjectures 
are in agreement with insights asserting that such 
statistical de-noising methods will leave
a substantial amount of noise uncorrected. Moreover, ``natural'' examples of 
noisy highly correlated classical systems exhibit a moderate 
degree of dependence, much 
less than the sort of dependence required 
for quantum error-correction, and appear to be in agreement 
with Conjecture [C]. 


\section {The nature of noise and the rate of errors}
\label {s:rate}

Up to now, we have assumed
that the error rate (per qubit, in each computer cycle) is small 
and fixed. Trying 
to understand systematic relations 
between the error rate (for individual qubits)  
and the intended state $\rho$ of the quantum computer 
may be of interest. 

In this section we assume that an evolution of a 
noisy quantum computer (or a more general noisy quantum system) is described 
for a certain time interval. The intended state of the system is 
pure for the entire time, and we assume, as before, that for the entire time, 
the actual state of the 
computer is close to the intended state. Up to 
now, we took a ``snapshot'' at a single computer cycle and measured 
the error-rate for individual qubits. Here we will consider the
infinitesimal noise rate in terms of trace distance.

\begin {itemize}

\item [{\bf [E.1]}]
A noisy quantum system 
is subject to (detrimental) noise with the following property: 
the infinitesimal rate of noise at time $t$ (in terms of trace distance) is 
bounded from below by a measure of 
noncommutativity 
between 
the operators describing the
evolution 
prior to time $t$ and those describing it after time $t$.  
\end {itemize}


Conjecture [E.1]
is an attempt to describe in a more concrete 
form the postulate of noise itself. 
Conjecture [E.1] 
seems similar to  
models of decoherence (brought to my attention by 
Michael Khasin) 
where  the decoherence is described as the effect of several 
noncommuting noise operations, and the rate of decoherence is related 
to ``uncertainty measures'' for   
these noise operators. 

When we consider a quantum computer with $n$ qubits 
whose evolution is restricted to a Hilbert space $V$ (which can be a subspace
of the entire $2^n$-dimensional space of pure states for all the qubits) 
we can expect that the rate of noise (in terms of trace distance) 
will be bounded below by $\kappa \log (\dim V)$. 
This appears to be in agreement with the behavior for unprotected noisy 
quantum circuits. 

\begin {itemize}
\item [{\bf [E.2]}]
A noisy quantum computer with $n$ (logical) qubits 
is subject to (detrimental) noise, with the following property: 
For some $\kappa>0$, 
the infinitesimal rate of noise 
(in terms of trace distance)  
is at least $\kappa n$.  

\end {itemize}

For standard models of noise, the infinitesimal rate of noise 
in terms of trace distance scales up linearly with the number of qubits. 
So Conjecture [E.2] 
is in agreement with the standard assumptions 
on the rate of noise. (The rate of noise is a small constant 
per qubit per one computer cycle). 
However, there are two important differences.

1) For strong forms of error-synchronization (like the model 
in Section \ref {s:abqec}) the amount of error in terms of 
trace distance, is sublinear in the number of qubits.  
Therefore,  in cases of highly entangled
states, which, by Conjecture [B], would lead to error-synchronization,  
conjecture [E.2] suggests that the rate of detrimental 
noise for individual qubits 
will scale up (even linearly) with the number of involved qubits.

2) Ordinary models of noise enable the existence of 
``decoherence-free subspaces.'' Conjecture [E] asserts that, in contrast, 
the (detrimental) decoherence rate for a subspace representing a 
small number of ``protected logical qubits'' is ``intrinsic,'' depends 
on the space of 
operators acting on this subspace along the evolution of the 
computer, and, in term of trace-distance, scale up linearly 
with the number of qubits.

{\bf Remark: } When we wish to prescribe an evolution of a quantum 
system up to a small error, a lower bound on the error rate at an 
intermediate state $\rho$ may depend 
not only on the process leading to $\rho$ from the initial state 
but also on the process leading 
from it to the terminal state. For example, we can expect that the 
noise for two faraway entangled photons will be independent, and therefore
the rate of detrimental decoherence in this case is zero.
Note that this is consistent with Conjecture [E.1]. 
Any intervention to bring the two photons 
back together in order to carry out additional 
joint operations is expected to introduce 
strong correlation between their errors. 

More generally, the point is this: 
Consider an intended pure-state evolution $\rho_t$, $0\le t\le 1$ of 
a quantum computer, and a noisy realization $\sigma_t$, $0 \le t \le 1$. 
Assuming that $\sigma$ is close to $\rho$ for the {\it entire} 
time interval, may create dependence of the infinitesimal noise at an 
intermidiate time $t$ on the entire evolution of $\rho$.       


 


\section {Conclusion}
\label {s:con}

If (or when) true, our  
conjectures on the nature of information leaks (decoherence)
for quantum computers
are damaging to the possibility of storing and
manipulating highly entangled quantum qubits. 
The conjectures do not contravene quantum mechanics nor, to the 
best of my knowledge, 
established physics phenomena. 
Neither do our conjectures contravene the feasibility of classical forms 
of error-correction and fault-tolerant computation.

Testing these conjectures empirically may be possible
for quantum computers with a relatively small number of qubits.  
The conjectures might also be refuted by constructions of 
highly stable qubits based on strong entanglement, such as
stable non-Abelian anyons 
\cite {K2,FKLW,MR}.



\section {Appendix}

\subsection {Another measure for information leaks}

Given a quantum operation $E$, our measure $$L(A)=L_E(A;\tau)$$ 
for the information leaks 
for a set $A$ of qubits depended on a pure tensor product state $\tau$.
The two-qubits basic property of detrimental decoherence was made 
for every $\tau$ separately. For the stronger conjectures below we will continue to make the 
statements in terms of an auxiliary tensor product state $\tau$. 
We will write
$L(A)=L_E(A;\tau)$ and similarly delete $E$ and $\tau$ from 
other definitions based on $L(A)$.

An alternative approach is as follows. 
Let $\psi$ be the state of the computer's qubits and the 
environment that is represented by a set $N$ of qubits. Let $U$ be a unitary operator
of the computer and environment qubits representing the noise. A standard measure
of the information that the environment has on the qubits in $A$  
is $$L'(A) = S(U (\psi) |_A) 
+ S(U( \psi) |_N) - S(U( \psi)|_{A\cup N }).$$
For our purposes we take $\psi = \psi_0 (A) \otimes \psi_1(N)$ 
where $\psi_1(N)$ is any pure state on the environment and $\psi_0(A)$
is the mixed state of maximum entropy on $A$.
I would expect that $L'(A)$ can replace $L(A)$ for the formulation 
of Conjecture [A] and the stronger conjectures below. 




\subsection {More qubits}

Here is a suggestion for an extension of the above 
conjecture from pairs of qubits to  larger sets of qubits. 
This suggestion goes beyond Conjectures [A] and [B] 
and is related to strong  
error synchronization.


For a set $A = \{a_1,a_2,\dots,a_m\} $ of $m$ qubits recall that

$$ENT(\rho; A) = -S(\rho)+ \max S(\rho^*),$$ 
where $\rho^*$ is a mixed state with the same marginals
on proper sets of qubits as $\rho$, i.e., 
$\rho^*|_B = \rho|_B$ for every proper subset $B$ of $A$.

Define in a similar way $$EL(A) = -L_E(A)+\max L_{E^*}(A),$$
where $E^*$ is a quantum operation that satisfies
$E^*|_B=E|_B$ for every proper set $B$ of $A$.

Using these definitions we will extend our conjectures, given by 
relations (\ref {e:p1}) and (\ref {e:p1strong}), from pairs of qubits 
to larger sets of qubits. 
Let $\rho$ be an ideal state of the computer and 
let $A$ be a set of $m$ qubits. Extending (\ref{e:p1}) we conjecture that

\begin {equation}
\label {e:p1s}
EL(A) 
\ge  
K_m 
ENT(\rho|_A). 
\end {equation}

\noindent
Here, $K_m = K_m(\{L(a): a \in A\})$ is substantially 
larger than $\min \{L(a))):a \in A\}^2$ and it vanishes when all the 
individual information leaks vanish. 

Here again we  
further conjecture that for every representation $\omega$ 
of the state $\rho|_A$ 
as a convex combination $\rho|_A = \sum p_k \rho_k$ of pure
joint states,   

\begin {equation}
\label {e:p1sstrong}
EL(A) 
\ge  
K_m 
\sum p_k ENT(\rho_k; A). 
\end {equation}


{\bf Remark:} The value of $ENT(\rho; A)$ is intended to 
serve as a measure of the additional information when we pass 
from ``marginal distributions'' on proper subsets of 
qubits to the entire distribution on all qubits. 

The additional conjectures of this section 
are meant to draw the following picture: 
we have an ideal notion of a quantum computer that has 
extraordinary physical and computational properties.
Next come noisy quantum computers with an ideal notion of noise.
If the noise rate is small then FTQC is possible.
Next come noisy quantum computers that satisfy 
relation (\ref {e:p1}). For them, 
fault-tolerance will require controlling the error rate as well as
$K_2$, which we expect to be much harder. This model
is also an idealization as long as $K_3=0$ and so on. For such 
highly entangled
states as those required in quantum algorithms, (\ref{e:p1sstrong}) will be 
more and more damaging for larger values of $i$.





\subsection {Mathematical challenges}

We will mention now some mathematical challenges.
It will be interesting to prove relation (\ref {e:c}) based on
relation (\ref {e:p1s}), and to formulate and prove 
weak and strong forms of 
error synchronization based, respectively, on 
relations (\ref {e:p1}) and (\ref {e:p1s}).
A further goal would be to derive, based on the assumptions 
on noise for the physical qubits 
(relations (\ref{e:p1s}) and  (\ref{e:p1sstrong})), the same relations
as well as relation (\ref {e:c})  
for  ``protected'' qubits, namely
logical qubits represented by quantum error-correction. It will also 
be of interest to find the right general 
formulation of ``tend to commute'' as in relation (\ref{e:t2c}) 
and to relate it to the more concrete conjectures for quantum computers.







\end {document}